\documentclass[preprint,12pt]{elsarticle}

\usepackage{amssymb}

\journal{Mathematics and Computers in Simulation}

\begin{document}

\begin{frontmatter}


\title{On some special solutions to 
periodic Benjamin-Ono equation with discrete Laplacian}

\author{Yohei Tutiya \\
(Ohara graduate school of accounting, Tokyo, Japan)\\
Jun'ichi Shiraishi\\
(Graduate School of Mathematical Sciences, Tokyo, Japan)}




\begin{abstract}
We investigate a periodic version of the Benjamin-Ono (BO)
equation associated with a discrete Laplacian. 
We find some special solutions to this equation, and 
calculate the values of the first two integrals of motion 
$I_1$ and $I_2$ corresponding to these solutions.
It is found that there exists a strong resemblance between 
them and the spectra for the Macdonald $q$-difference operators.
To better understand the connection between these classical and quantum integrable systems, we consider 
the special degenerate case corresponding to $q=0$ in more detail.
Namely, we give general solutions to this degenerate periodic BO,
obtain explicit formulas representing all the integrals of motions $I_n$ ($n=1,2,\cdots$),
and successfully identify it with the eigenvalues of 
Macdonald operators in the limit $q\rightarrow 0$, i.e. the limit where 
Macdonald polynomials tend to the Hall-Littlewood polynomials.

\end{abstract}


\end{frontmatter}


\section{Introduction}
\label{introduction}

In \cite{ST}, the authors studied 
a doubly periodic version of the 
intermediate long wave (ILW) equation associated 
with a discrete Laplacian. Then this ILW-type equation was
identified with the first member of an integrable hierarchy 
which is given as a certain reduction of the discrete KP theory 
in the framework of the Sato theory.

In this paper,  we study a spacial case 
having a single periodicity, instead of dealing with the most general doubly periodic case
which requires much more technique based on the algebro-geometric argument.
Namely we consider the periodic Benjamin-Ono equation \cite{B,O} 
with discrete Laplacian
\begin{eqnarray}
&&\frac\partial{\partial t}\eta(x,t)
=
\eta(x,t)\int_{-1/2}^{1/2}
\hspace{-2.2em}\backslash\hspace{.5em}\,\,\,\,\,
\left[\cot\big\{\pi(y-x-\gamma)\big\}
\right.\nonumber\\
&&\hspace{5em}\left.-2\cot\big\{\pi(y-x)\big\}
+\cot\big\{\pi(y-x+\gamma)\big\}\right]
\eta(y,t)\frac{idy}{2},
\label{KotenMacd}
\end{eqnarray}
where the symbol
$\int\hspace{-0.8em}\backslash\hspace{.5em}$
denotes the Cauchy principal value integral and 
$\gamma$ is a complex constant with 
nonzero imaginary part.
Our goal in this paper is to study some special solutions to (\ref{KotenMacd}) and
the properties of the associated integrals of motion in some detail.
We also investigate the special case $\gamma\rightarrow i \infty$ in great detail since 
it allows us to have very much explicit results about solutions and integrals.
\bigskip

We claim that (\ref{KotenMacd}) 
can be regarded as a classical integrable system
associated with the Macdonald theory of symmetric functions
with two parameters $q$ and $t$ \cite{Mac}. 
Recall that 
Macdonald introduced a set of commuting $q$-difference operators
$D_1,D_2,\cdots$ depending on $q$ and $t$, which are acting
 on the space of symmetric polynomials, say in $x_i$'s.
Then he proved the fundamental existence theorem for
the simultaneous eigenfunctions 
$P_\lambda(x;q,t)$ of the $D_i$'s, together with a certain normalization condition. 
Here the index $\lambda=(\lambda_1,\lambda_2,\cdots)$
denotes the partition. 

For our purpose to compare (\ref{KotenMacd}) with 
the Macdonald theory, we better consider the 
infinitely many variable case and use the Heisenberg 
representation of the Macdonald operators developed
in [AMOS], [S] and [FHHSY].
Introduce the Heisenberg algebra given by the generators
$\{a_n\}_{n\in {\bf Z}_{\neq 0}}$ and the commutation relations
\begin{eqnarray}
[a_m, a_n]=m\frac{1-q^{|m|}}{1-t^{|m|}}
\delta_{m+n,0}.
\end{eqnarray}
Let $\varepsilon$ be a constant. Set
\begin{eqnarray}
\eta(z)=\sum_{n \in {\bf Z}} \eta_{n}z^{-n}=\varepsilon 
\exp\left(\sum_{n>0}{1-t^{-n}\over n}a_{-n}z^n\right)
\exp\left(-\sum_{n>0}{1-t^{n}\over n}a_{n}z^{-n}\right).
\label{vertex}
\end{eqnarray}
It was shown in [S] and [FHHSY] that we have an infinite commutative family of 
operators acting on the Fock space containing $\eta_0$ as the first nontrivial member. 

One of the canonical ways to write down the generators of this commutative algebra
is as follows. Set
\begin{eqnarray}
I_n=\oint\cdots \oint_C
{dz_1\over 2 \pi i z_1} \cdots {dz_n\over 2 \pi i z_n} 
\prod_{1\leq j<k\leq n}
{z_j-z_k\over z_j-q z_k}
\cdot :\eta(z_1)\cdots \eta(z_n):,
\label{In}\end{eqnarray}
where the integration contour $C$ is the torus $|z_j|=1,j=1,\cdots,n$,
and the symbol $:\cdots :$ denotes the ordinary normal ordering of the 
Heisenberg generators, more explicitly we have
\begin{eqnarray}
:\eta(z_1)\cdots \eta(z_n):=
\prod_{1\leq j<k\leq n}
{(z_j-qz_k)(z_j- t^{-1}z_k)\over (z_j- z_k)(z_j-qt^{-1} z_k)}\cdot \eta(z_1)\cdots \eta(z_n).
\end{eqnarray}
Then we have $[I_m,I_n]=0$ for all $m,n\geq 1$. Moreover 
we can write down explicitly the spectrum of $I_n$ 
as $q^{-n(n-1)/2}\prod_{k=1}^n (1-q^k)\cdot e_n(\varepsilon t^{-\lambda_1},
\varepsilon q t^{-\lambda_2},\varepsilon q^2 t^{-\lambda_3},\cdots)$ 
where $e_n(x_1,x_2,\cdots)$ 
denotes the $n$-th elementary symmetric function,
$\lambda=(\lambda_1,\lambda_2,\cdots)$ is a partition.

For simplicity, we set $\alpha_n=-(1-t^n)a_n/n$. With this notation, we have 
$\eta(z)=\varepsilon  :\exp\left(\sum_{n\neq 0}\alpha_n z^{-n} \right):$.
Now we proceed to considering a classical limit. 
Set $t=e^\hbar$. While fixing $q$, we consider the limit $\hbar\rightarrow 0$, namely $t\rightarrow 1$.
In this limit we have $[\alpha_m,\alpha_n]= 0$ and $[\eta_m,\eta_n]=0$.
Hence we regard
the algebra generated by $\alpha_n$'s or
$\eta_n$'s being our algebra of classical observables.

Induce the poisson bracket by $\{u,v\}:=\lim_{\hbar\to0}[u,v]/\hbar$ as usual.
Then we have the set of canonical commutation relations
$\{\alpha_m,\alpha_n\}={\rm sgn}(m)(1-q^{|m|})\delta_{m+n,0}$, which gives us
$\{\eta_m,\eta_n\}=\sum_{l\neq 0}{\rm sgn}(l)(1-q^{|l|})\eta_{m-l}\eta_{n+l}.$

Let $\eta_0$ be our Hamiltonian.
For example, the time evolution of $\eta(z)$ is given by
\begin{eqnarray}
{\partial \over \partial t}\eta(z)=\{\eta_0, \eta(z)\}
=\eta(z)\sum_{n\neq 0} {\rm sgn}(n)(1-q^{|n|})\eta_{-n}z^n,
\label{fourier}\end{eqnarray}
which is just identical to (\ref{KotenMacd})
under the identification of the variables $z=\exp(2\pi i x)$ and $q=\exp(2\pi i \gamma)$.

Set 
\begin{eqnarray}
\tau_+(z)=\exp\left(-\sum_{n>0}{1\over 1-q^{n}}\alpha_{-n}z^n\right),
\tau_-(z)=\exp\left(-\sum_{n>0}{1\over 1-q^{-n}}\alpha_{n}z^{-n}\right).
\end{eqnarray}
We have
\begin{eqnarray}
&&\eta(z)=\varepsilon {\tau_-(q^{-1}z)\over \tau_-(z)}{\tau_+(qz)\over \tau_+(z)},\\
&&D_t \;\; \tau_-(z)\cdot \tau_+(z)=\varepsilon \tau_-(q^{-1}z)\tau_+(qz)-
\eta_0 \tau_-(z)\tau_+(z).
\label{bilinear}\end{eqnarray}

Here comes a question: 
how one can find some special solutions to (\ref{KotenMacd}) or (\ref{bilinear}),
calculate integrals of motion, and find some connection
with the theory of Macdonald symmetric polynomials? In this paper,
we show some 
explicit connection between the (classical) integrals and the (quantum) eigenvalues,
indicating that some deep structure is hidden in our problem.
\bigskip

This paper is organized as follows.
In \S2, we give some special solutions to (\ref{KotenMacd}).\
The first few $I_n$'s of them are calculated explicitly. It is shown that
they have precisely the same formulas as the ones for the eigenvalues of the Macdonald operators.\ 
In \S3, the special case $q=0$ is treated in a complete manner. 
The general formula for $I_n$'s is given.

A remark is in order here.
In the work by Avanov, Bettelheim and Wiegmann \cite{ABW},
a classical integrable system associated with the Calogero-Sutherland model
is studied.
It is an interesting future problem to connect their bilinear equation and solutions 
with ours.

\section{Some special solutions to the periodic BO equation (\ref{KotenMacd})}
After a little algebra, one can construct some Laurent polynomial solutions to 
the bilinear equation  (\ref{bilinear}).
Recall that $\varepsilon$ is a parameter in (\ref{bilinear}), and 
we will consider it as an arbitrary parameter.
We have a class of solutions parametrized by one more parameter $a$.
One may easily find $\tau_+$ and $\tau_-$ given by 
\begin{eqnarray}
&&\left\{\begin{array}{l}
\displaystyle\tau_+=1+ze^{(1-q)at},\\
\displaystyle\tau_-=1+
\frac{\varepsilon-a}{\varepsilon-qa}
z^{-1}e^{-(1-q)at},
\end{array}\right.\label{g1solution}
\end{eqnarray}
satisfy (\ref{bilinear})
under the condition that $\varepsilon q+(1-q)a=\eta_0=I_1$.

We can compute higher integrals of motion (\ref{In}) corresponding to this special solution.
For example, we have
\begin{eqnarray}
\begin{array}{l}
\displaystyle
I_1=\varepsilon q+(1-q)a
=(1-q)e_1(a,\varepsilon q,\varepsilon q^2,\cdots),\\[.5em]
\displaystyle
I_2=\varepsilon^2q^2+(1-q^2)a\varepsilon
={(1-q)(1-q^2)\over q}e_2(a,\varepsilon q,\varepsilon q^2,\cdots),\  
\end{array}
\end{eqnarray}
and so on.

Next, we show another class of special solutions to (\ref{bilinear})
which has two parameters $a_1$ and $a_2$. Set
\begin{eqnarray}
&&\left\{\begin{array}{l}
\displaystyle\tau_+=
1+ze^{(1-q)a_1t}+ze^{(1-q)a_2t}\\
\displaystyle\hspace{6em}
+{(a_1-a_2)^2\over(a_1-qa_2)(a_1-q^{-1}a_2)}
z^2e^{(1-q)(a_1+a_2)t},\\[1.2em]
\displaystyle\tau_-=1+
c_1z^{-1}e^{-(1-q)a_1t}
+c_2z^{-1}e^{-(1-q)a_2t}\\
\displaystyle\hspace{6em}+
{(a_1-a_2)^2\over(a_1-qa_2)(a_1-q^{-1}a_2)}
c_1c_2z^{-2}e^{-(1-q)(a_1+a_2)t},
\end{array}\right.\label{g2solution}\\[.3em]
&&c_j=\frac{\varepsilon-qa_j}{\varepsilon-q^2a_j} {(a_1-qa_2)(a_1-q^{-1}a_2)\over (a_1-a_2)^2}.
\end{eqnarray}
Then the (\ref{bilinear}) holds under the condition that 
$\varepsilon q^2+(1-q)a_1+(1-q)a_2=\eta_0=I_1$. 
One can calculate the higher integrals of motion (\ref{In}). 
The first two read
\begin{eqnarray}
\begin{array}{l}
\displaystyle
I_1=\varepsilon q^2+(1-q)a_1+(1-q)a_2
=(1-q)e_1(a_1,a_2,\varepsilon q^2,\varepsilon q^3,\cdots),\\[.5em]
\displaystyle
I_2={(1-q)(1-q^2)\over q} a_1a_2+\varepsilon q(1-q)a_1
+\varepsilon q(1-q)a_2+\varepsilon^2q^2\\[.9em]
\displaystyle\hspace{1em}={(1-q)(1-q^2)\over q}e_2(a_1,a_2,\varepsilon q^2,\varepsilon q^3,\cdots),\  
\end{array}
\end{eqnarray}
and so on. 

It is clearly seen from these examples that there 
is a strong resemblance between such formulas for the integrals of motion and 
the eigenvalues of the Macdonald operators 
$q^{-n(n-1)/2}\prod_{k=1}^n (1-q^k)\cdot e_n(\varepsilon t^{-\lambda_1},
\varepsilon q t^{-\lambda_2},\varepsilon q^2 t^{-\lambda_3},\cdots)$.
Namely, if we take the limit $t\rightarrow 1$, $\lambda_i\rightarrow \infty $ 
in such a way that we have the finite limits $\lim \varepsilon q^{i-1}t^{-\lambda_i}=a_i$,
then we recover the above formulas for the integrals of motion.
At present the reason of this beautiful correspondence has not been investigated.

\section{the spacial case $q=0$}
At present, unfortunately, 
our study on the special solutions to (\ref{bilinear}) and whose integrals of motion
still remains primitive and heuristic. 
In the special limit $q=0$, however, we can easily complete our 
program within the technique of linear algebra as we will show below.

Before we embark on the study of the bilinear equation,
let us try to solve naively  (\ref{fourier}) in the limit $q=0$, namely
\begin{eqnarray}
{\partial \over \partial t}\eta(z)
=\eta(z)\sum_{n\neq 0} {\rm sgn}(n)\eta_{-n}z^n.
\end{eqnarray}
Set 
$\eta_\pm(z):=\sum_{n>0}\eta_{\mp n}z^{\pm n}$. Then we have
the split equations
\begin{eqnarray}
{d\over dt}\eta_+(z) =\eta_+(z)(\eta_+(z)+\eta_0),
\quad {d\over dt}\eta_-(z) =-\eta_-(z)(\eta_-(z)+\eta_0).
\end{eqnarray}
One immediately finds the general solution, 
\begin{eqnarray}
\eta_+(z)={-\eta_0 c_+(z) e^{\eta_0 t}\over d_+(z)+c_+(z) e^{\eta_0 t}},
\qquad \eta_-(z)={-\eta_0 c_-(z) e^{-\eta_0 t}\over d_-(z)+c_-(z) e^{-\eta_0 t}}
\label{tau-eta}
\end{eqnarray}
where $d_\pm(z),c_\pm(z)$ are arbitrary holomorphic functions in 
some neighborhoods of $z^{\pm 1}=0$.

One may ask if  $d_\pm(z),c_\pm(z)$ can be restricted to Laurent polynomials in $z^{\pm 1}$,
and the denominators $d_\pm(z)+c_\pm(z) \exp(\pm \eta_0 t)$ play the role of
the tau functions $\tau_\pm(z)$.
The answer indeed is yes.

In the case $q=0$, we have 
\begin{eqnarray}
&&\eta(z)=\varepsilon {1\over \tau_-(z) \tau_+(z)},\label{eta-tau}\\
&&D_t \;\; \tau_-(z)\cdot \tau_+(z)=\varepsilon -
\eta_0 \tau_-(z)\tau_+(z).
\label{bilinear-q=0}\end{eqnarray}
Setting $\tau_\pm(z)=d_\pm(z)+c_\pm(z) \exp(\pm\eta_0 t)$, one finds that 
(\ref{bilinear-q=0}) reduces to the functional equation
\begin{eqnarray}
d_+(z)d_-(z)-c_+(z)c_-(z)=\varepsilon/\eta_0.
\label{dc}\end{eqnarray}

\noindent{\bf Proposition.}
Let $n$ be a nonnegative integer, and let $\epsilon_{\pm1},\cdots,\epsilon_{\pm n}$
be $2n$ parameters. 
Let $e_{m}(z,t)=\epsilon_m z^m\exp\{{\rm sgn}(m)\eta_0 t\}$ for $m\in {\bf Z}_{\neq 0}$, and define
$\tau_\pm(z,t;\epsilon_{\pm 1},\cdots,\epsilon_{\pm n})=\tau_\pm(z)$ by
\begin{eqnarray}
\tau_+(z)&=&1+\sum_{1\leq i\leq n} e_i+
\sum_{1\leq i_1<i_2\leq n} e_{i_2}e_{-i_1}+
\sum_{1\leq i_1<i_2<i_3\leq n} e_{i_3}e_{-i_2}e_{i_1}+\nonumber\\
&&+\cdots +e_{n}e_{-n+1}e_{n-2}\cdots e_{-(-1)^n},\label{tauplus}\\
\tau_-(z)&=&1+\sum_{1\leq i\leq n} e_{-i}+
\sum_{1\leq i_1<i_2\leq n} e_{-i_2}e_{i_1}+
\sum_{1\leq i_1<i_2<i_3\leq n} e_{-i_3}e_{i_2}e_{-i_1}+\nonumber\\
&&+\cdots +e_{-n}e_{n-1}e_{-n+2}\cdots e_{(-1)^n}.\label{tauminus}
\end{eqnarray}
Then these satisfy (\ref{bilinear-q=0})
under the condition that $\prod_{i=1}^n (1-\epsilon_i\epsilon_{-i})=\varepsilon/\eta_0$.

\noindent{\bf Proof.} 
We have
\begin{eqnarray}
&&d_+(z)=1+\sum_{1\leq i_1<i_2\leq n} \epsilon_{i_2}\epsilon_{-i_1}z^{i_2-i_1}+\cdots,\\
&&c_+(z)=\sum_{1\leq i\leq n} \epsilon_i z^i+
\sum_{1\leq i_1<i_2<i_3\leq n} \epsilon_{i_3}\epsilon_{-i_2}\epsilon_{i_1}z^{i_3-i_2+i_1}+\cdots,\\
&&d_-(z)=1+\sum_{1\leq i_1<i_2\leq n} \epsilon_{-i_2}\epsilon_{i_1}z^{-i_2+i_1}+\cdots,\\
&&c_-(z)=\sum_{1\leq i\leq n} \epsilon_{-i} z^{-i}+
\sum_{1\leq i_1<i_2<i_3\leq n} \epsilon_{-i_3}\epsilon_{i_2}\epsilon_{-i_1}z^{-i_3+i_2-i_1}+\cdots,
\end{eqnarray}
and need to show that
\begin{eqnarray}
d_+(z)d_-(z)-c_+(z)c_-(z)=\prod_{i=1}^n (1-\epsilon_i\epsilon_{-i}).
\label{cc}\end{eqnarray}
A way to write $d_\pm$ and $c_\pm$ is
\begin{eqnarray}
&&\left(\begin{array}{cc}
\displaystyle d_{+}(z)&c_{+}(z)\\[.3em]
\displaystyle c_{-}(z)&d_{-}(z)
\end{array}\right)
=
\left(\begin{array}{cc}
\displaystyle1&\epsilon_{n}z^n\\
\displaystyle \epsilon_{-n}z^{-n}&1
\end{array}\right)
\cdots
\left(\begin{array}{cc}
\displaystyle1&\epsilon_{1}z\\
\displaystyle \epsilon_{-1}z^{-1}&1
\end{array}\right).
\end{eqnarray}
The determinant of the above gives the relation (\ref{cc}).\qed
\medskip

\noindent{\bf Remark.}
For $n\geq0$, and $\epsilon_{\pm 1},\cdots,\epsilon_{\pm n}$,
we have constructed a solution to (\ref{bilinear-q=0})
and have denoted them by $\tau_\pm(z,t;\epsilon_{\pm 1},\cdots,\epsilon_{\pm n})$.
It is desirable, however, for our later purpose to have more 
flexible notation in which we can treat $\tau_\pm$ with different numbers of parameters 
within a unified notation.
It is clear that from infinite sequences $(\epsilon_{\pm 1},\epsilon_{\pm 2},\cdots)$
with only finitely many nonzero parts, say $\epsilon_{\pm i}=0$ for $i>n$, 
we can construct the tau functions as above.
We will denote them by $\tau_\pm(z,t;\epsilon_{\pm 1},\epsilon_{\pm 2},\cdots)$ 
without the need of referring to the `$n$'.

\vspace{\baselineskip}
When $q=0$, the integral of motion (\ref{In}) becomes the Toeplitz determinant
\begin{eqnarray}
I_{n}=
\left|\begin{array}{cccc}
\displaystyle\eta_0&\eta_{-1}&\cdots&\eta_{-n+1}\\
\displaystyle\eta_{1}&\eta_0&\cdots&\eta_{-n+2}\\
\vdots&\vdots&\ddots&\vdots\\
\displaystyle\eta_{n-1}&\eta_{n-2}&\cdots&\eta_0\\
\end{array}\right|.
\end{eqnarray}

\noindent{\bf Theorem.}
Let $\eta_0$ be satisfying the condition
$\prod_{i=1}^\infty (1-\epsilon_i\epsilon_{-i})=\varepsilon/\eta_0$, and 
let $\tau_\pm(z)$ be defined by (\ref{tauplus}) and (\ref{tauminus}).
Hence (\ref{bilinear-q=0}) is satisfied. 
Then define $\eta_n$ by (\ref{eta-tau}) or (\ref{tau-eta}). 
Then we have 
\begin{eqnarray}
I_{k+1}=\eta_0^{k+1}(1-\epsilon_1\epsilon_{-1})^k
(1-\epsilon_2\epsilon_{-2})^{k-1}
\cdots(1-\epsilon_k\epsilon_{-k}),
\end{eqnarray}
for $k\geq 1$.

\noindent
{\bf Proof.}
We know that $I_n$'s are $t$ independent. Thus we may set $t=0$ in what follows
in order to make our computation simple. 
Set
\begin{eqnarray}
\eta_{\pm}(z,0)={-\eta_0c_{\pm}(z)\over d_{\pm}(z)+c_{\pm}(z)}
=:-\eta_0\sum_{n>0}{\nu_{\pm n}}z^{\pm n},\
\end{eqnarray}
for notational simplicity.
Then, our claim in Theorem is 
\begin{eqnarray}
\left|\begin{array}{cccc}
\displaystyle-1&\nu_{-1}&\cdots&\nu_{-k}\\
\displaystyle\nu_{1}&-1&\cdots&\nu_{-k+1}\\
\vdots&\vdots&\ddots&\vdots\\
\displaystyle\nu_{k}&\nu_{k-1}&\cdots&-1\\
\end{array}\right|
=(-1)^{k+1}\prod_{l=1}^{k}(1-\epsilon_l\epsilon_{-l})^{k-l+1}. \label{mochi}
\end{eqnarray}
This can be proved by Lemma stated below as
\begin{eqnarray}
&&I_{k+1}=
\left|\left(\begin{array}{cccc}
\displaystyle-1&\nu_{-1}&\cdots&\nu_{-k}\\
\displaystyle\nu_{1}&-1&\cdots&\nu_{-k+1}\\
\vdots&\vdots&\ddots&\vdots\\
\displaystyle\nu_{k}&\nu_{k-1}&\cdots&-1\\
\end{array}\right)
\left(\begin{array}{cccc}
1&0&\cdots&0\\
\tau_1^{(k)}&1&0\cdots&0\\
\vdots&\vdots&\ddots&\vdots\\
\tau_k^{(k)}&0&\cdots&1
\end{array}\right)\right|\nonumber\\
&&\quad=\left|\begin{array}{cccc}
\displaystyle-(1-\epsilon_1\epsilon_{-1})
\cdots(1-\epsilon_k\epsilon_{-k})
&\nu_{-1}&\cdots&\nu_{-k}\\
\displaystyle0&-1&\cdots&\nu_{-k+1}\\
\vdots&\vdots&\ddots&\vdots\\
\displaystyle0&\nu_{k-1}&\cdots&-1\\
\end{array}\right|\nonumber\\[.5em]
&&=-(1-\epsilon_1\epsilon_{-1})
\cdots(1-\epsilon_k\epsilon_{-k})I_{k}.
\nonumber\end{eqnarray}
\qed

\noindent{\bf Lemma.} For $N\geq 1$, we have
\begin{eqnarray}
\left(\begin{array}{cccc}
\displaystyle-1&\nu_{-1}&\cdots&\nu_{-N}\\
\displaystyle\nu_{1}&-1&\cdots&\nu_{-(N-1)}\\
\vdots&\vdots&\ddots&\vdots\\
\displaystyle\nu_{N}&\nu_{N-1}&\cdots&-1\\
\end{array}\right)
\left(\begin{array}{c}
1\\
\tau_1^{(N)}\\
\vdots\\
\tau_N^{(N)}
\end{array}\right)
=
\left(\begin{array}{c}
-(1-\epsilon_1\epsilon_{-1})
\cdots(1-\epsilon_N\epsilon_{-N})\\
0\\
\vdots\\
0
\end{array}\right).
\nonumber\end{eqnarray}
Here, $\tau_{\pm j}^{(m)}$ is the coefficient of $z^{\pm j}$ in 
$\tau_\pm(z,0;\epsilon_{\pm 1},\epsilon_{\pm 2},\cdots,\epsilon_{\pm m},0,0,\cdots)$, namely
\begin{eqnarray}
\tau_\pm(z,0;\epsilon_{\pm 1},\epsilon_{\pm 2},\cdots,\epsilon_{\pm m},0,0,\cdots)
=\sum_{j=1}^m\tau_{\pm j}^{(m)}z^{\pm j}.
\end{eqnarray}

\noindent{\bf Sketch of the proof of Lemma.}
We have the following three
properties,
which can be proved by induction on $N$. 

\begin{itemize}
\item[1.]
$\nu_N$ is a polynomial depending only on $\epsilon_{\pm1},
\epsilon_{\pm2},\cdots,\epsilon_{\pm (N-1)}$ and 
$\epsilon_{N}$, and 
$\nu_{-N}$ is a polynomial depending only on $\epsilon_{\pm1},
\epsilon_{\pm2},\cdots,\epsilon_{\pm (N-1)}$ and 
$\epsilon_{-N}$.
\item[2.]
$\displaystyle\frac{\partial \nu_{\pm N}}{
\partial \epsilon_{\pm N}}=
(1-\epsilon_1\epsilon_{-1})(1-\epsilon_2\epsilon_{-2})
\cdots(1-\epsilon_{N-1}\epsilon_{-N+1})
$.
\item[3.] We have
\begin{eqnarray}
&&\left(\begin{array}{cccc}
\displaystyle\tau_{-N}^{(N)}&0&\cdots&0\\
\displaystyle\tau_{-N+1}^{(N)}&\tau^{(N)}_{-N}
&\cdots&0\\
\vdots&\vdots&\ddots&\vdots\\
\displaystyle\displaystyle\tau_{-1}^{(N)}&
\displaystyle\tau_{-2}^{(N)}&\cdots&\displaystyle\tau_{-N}^{(N)}\\
\end{array}\right)
\left(\begin{array}{c}
\nu_1\\
\nu_2\\
\vdots\\
\nu_N
\end{array}\right)\nonumber\\[.3em]
&&\qquad=
\left(\begin{array}{c}
d_{-N+1}^{(N)}\\
d_{-N+2}^{(N)}\\
\vdots\\
d_{0}^{(N)}
\end{array}\right)
-
\left(\begin{array}{c}
0\\
0\\
\vdots\\
(1-\epsilon_1\epsilon_{-1})
\cdots(1-\epsilon_N\epsilon_{-N})
\end{array}\right),
\nonumber\end{eqnarray}
\end{itemize}
where, $d_{\pm j}^{(m)}$ is defined by
\begin{eqnarray}
d_\pm(z,0;\epsilon_{\pm 1},\cdots,\epsilon_{\pm m} ,0,0,\cdots)
=:\sum_{j=1}^md_{\pm j}^{(m)}z^j. \nonumber
\end{eqnarray}
\qed
\vspace{1cm}

\noindent{\bf Acknowledgement}
JS thanks M. Noumi, Y. Yamada, Y. Ohta, P. Wiegmann, O. Foda and B. Feigin 
for stimulating discussion.
YT is grateful to R. Willox for constant encouragement.







\end{document}